# Impact of aleatoric, stochastic and epistemic uncertainties on project cost contingency reserves

D. Curto, F. Acebes [*], J.M. González-Varona, D. Poza

*GIR INSISOC-Department of Business Organization and CIM, School of Industrial Engineering, University of Valladolid, Valladolid, 47011, Spain*

ABSTRACT

In construction projects, contingency reserves have traditionally been estimated based on a percentage of the total project cost, which is arbitrary and, thus, unreliable in practical cases. Monte Carlo simulation provides a more reliable estimation. However, works on this topic have focused exclusively on the effects of aleatoric uncertainty, but ignored the impacts of other uncertainty types. In this paper, we present a method to quantitatively determine project cost contingency reserves based on Monte Carlo Simulation that considers the impact of not only aleatoric uncertainty, but also of the effects of other uncertainty kinds (stochastic, epistemic) on the total project cost. The proposed method has been validated with a real-case construction project in Spain. The obtained results demonstrate that the approach will be helpful for construction Project Managers because the obtained cost contingency reserves are consistent with the actual uncertainty type that affects the risks identified in their projects.

## 1. Introduction

All projects involve risk because projects aim to create a unique outcome. In addition, projects are developed in an environment of ambiguity, complexity and uncertainty, which implies a lack of information and knowledge about the environment and the future (Baccarini, 1996; Cagliano et al., 2015; Fan et al., 2008; Lam and Siwingwa, 2017). This unstable, uncertain and changing environment is responsible for altering project objectives (Association for Project Management, 2010; Hosny et al., 2018; Seyedhoseini et al., 2009). To address the causes that influence project planning and its consequences, the Risk Management process has been developed over the years. It is a structured approach to address the implications of risk in projects and to establish contingencies that ensure development and meet the planned objectives.

Risk Management consists of three distinct phases: risk identification, assessment and response (Zhang and Fan, 2014). Identification corresponds to the process of identifying and documenting the risks that affect the project. Evaluation refers to the study of the identified risks by an analysis of characteristics and the occurrence probability and associated impact. Finally, risk response focuses not only on considering the appropriate steps to be taken to manage risks but also on implementing and assessing such responses.

During the risk assessment, two main types of analysis, namely qualitative and quantitative, are distinguished (Hong et al., 2016; Moret and Einstein, 2016). The former is the assessment of the priority of risks and their relevance in the project (Allahi et al., 2017; El-Sayegh, 2008; Gosling et al., 2013; Hosny et al., 2018; Moreno-Cabezali and Fernandez-Crehuet, 2020). This process is performed by studying the occurrence probability and the associated impact on scope, time and cost (Hillson, 2005; Moreno-Cabezali and Fernandez-Crehuet, 2020).

The quantitative assessment aims to numerically measure risk-related changes in project objectives (AACE - American Association of Cost Engineering, 2011; Kwon and Kang, 2019). Only those institutions with a high degree of maturity can incorporate a quantitative analysis into the standard risk management process to look beyond the project boundaries for elements that can alter the fundamental conditions of a project's planning (Cagliano et al., 2015). Precisely some techniques based on the quantitative assessment allow us to estimate the contingency margins for the cost and time buffers for the deadline (Long and Ohsato, 2008). It is a matter of establishing a reserve fund consisting of a time buffer and a reserve budget item to cover the impact of risks and uncertainty by protecting project owners from undesired results (Uzzafer, 2013).

Contingency needs to be adequately predicted, budgeted and controlled throughout the project implementation period and, therefore, many risk assessment techniques aim to estimate the minimum contingencies needed to ensure the project's success, and doing this in a structured approach to Risk Management involves a significant improvement in contingency estimation accuracy (Akintoye and MacLeod, 1997; Mak and Picken, 2000).

* Corresponding author.
*E-mail addresses:* david.curto@uva.es (D. Curto), fernando.acebes@uva.es (F. Acebes), josemanuel.gonzalez.varona@uva.es (J.M. González-Varona), poza@eii.uva.es (D. Poza).

Several authors have studied different processes to estimate and allocate contingencies in projects (Baccarini, 2004; Baccarini and Love, 2014; El-Kholy et al., 2020; Hammad et al., 2016; Idrus et al., 2011; Kwon and Kang, 2019; Lorance and Wendling, 2001; Thal et al., 2010; Touran, 2003; Traynor and Mahmoodian, 2019), but none has done this by conducting an integrated analysis of duration and cost risks to include all the possible uncertainty types that may impact project activities.

Elms (2004) and Frank (1999) distinguish between two uncertainty types: aleatoric uncertainty (described by variability, meaning that there is a wide range of possible outcomes) and epistemic uncertainty (due to ambiguity or imperfect knowledge). Hillson (2014) widens this division by adding two additional uncertainty types to the above classification: stochastic uncertainty (also called "event risk", and defined as "possible future events") and ontological uncertainty (also known as "unknown-unknowns", which is unknown knowledge of what is impossible to know). Chapman and Ward (2011) offer a similar classification by dividing uncertainty into four different types: ambiguity (lack of complete/perfect knowledge); inherent variability (implies the equivalent of the events that always occur); the uncertainty of events (involves events, conditions, circumstances or scenarios that may, or may not, happen and further associated specific responses); systemic uncertainty (involves simple forms of dependence or complex feedback and feeding relations, including general or systemic ones).

In this context, the present paper aims to propose a methodology to calculate cost contingencies and their allocation in the project. To do so, we use Monte Carlo Simulation (MCS), a quantitative risk management technique that is considered suitable for cost contingency allocation (Barraza and Bueno, 2007; Chang and Ko, 2017; Liu et al., 2017). According to this technique, the first step is to identify the project risks. Once they have been identified, the probability of, and impact, on the project's duration and cost objectives are estimated. Unlike previous works, in this paper we classify each risk according to the uncertainty type that causes it: aleatoric, stochastic, epistemic (Hillson, 2014). This classification allows us to model each risk according to a different distribution function depending on the uncertainty associated with each risk. After applying MCS, we obtain a cost distribution function that allows Project Managers to estimate the cost contingencies for their projects. The obtained cost distribution function depends on the uncertainty type that affects each identified risk.

Therefore, the novelty of our study, which differentiates it from other contingency reserve allocation techniques, is that simulation will include not only the aleatoric uncertainty of the project activities, but also other types of uncertainty, such as risk events (i.e., stochastic uncertainty) and epistemic uncertainty (i.e., imperfect knowledge of the risk). This will allow a contingency reserve consistent to be obtained with the specific type of uncertainty (i.e., aleatoric, stochastic or epistemic) associated with each risk identified in the project.

To validate the proposed method, we used data from a typical construction project in which an Experts Committee filled in a Risk Register, including all the uncertainties that can affect the project objectives (aleatoric, stochastic, and epistemic). After performing MCS and analysing the obtained results, the project's cost contingencies were determined. The contingency calculated by the method herein proposed offers a more reliable value (by considering the specific uncertainty type that affects each identified risk) *versus* the traditional estimate based on a simple percentage allocation on the base budget (used by the developer in this actual project). Finally, we demonstrate that the method proposed in this paper outperforms another related work that considers only the aleatoric uncertainty of activities in the quantitative analysis of contingencies, but simulation does not include any remaining uncertainties identified in the project.

From this point, the rest of the document is structured as follows. The literature review section reviews the most relevant research on the project cost contingencies analysis. Then we present the methodology followed throughout this research. In the next chapter, we apply the proposed method to a real construction project. We discuss the obtained results and compare them to other existing techniques featured in the literature, which allowed us to validate our proposal. Then the conclusions and contributions are summarised.

## 2. Literature review

Contingency reserves, an amount of funds added to the base cost estimate to cover the estimated uncertainty and risk exposure, are defined by Project Management Institute (2017) as the budget in the cost baseline that is allocated to identified risks. Contingency reserves are often viewed as part of the budget intended to address the known unknowns that can affect a project. At the beginning of the project in the planning phase, as data and information are lacking, it is necessary to estimate contingencies to correct any possible deviations of the project from its objectives (Kwon and Kang, 2019). Project Management Institute (2017) differentiates contingencies into two categories: contingency reserve for identified risks (known-unknowns); management reserve for unknown risks (unknown-unknowns). Both these risk types are handled differently. As known-unknown risks (known-unknowns) can be identified and, therefore, analysed and assessed, they can be integrated into a mathematical or simulation model that can be used to derive a quantitative contingency value for these known risks. For unknown-unknown risks (unknown-unknowns, i.e., risks that cannot be identified because we are not aware of their existence), only management reserves can be assigned to be treated. For known risks, we can numerically estimate the value of contingencies using one of the existing methods in the literature. For unknown risks, allocation of management reserves is based on experience or using a reserve percentage of the total budget.

The contingency estimate for known risks has traditionally been calculated as an arbitrary percentage of the budgeted base cost (Ahmad, 1992; Moselhi, 1997). This percentage contingency allocation method has been a widespread technique and is mostly followed by some organisations with little or no risk maturity (Cagliano et al., 2015). The employed percentage can be modified from one project to another, or even in the same project. In other cases, this allocation percentage is established according to the value of the occurrence probability of risks based on the cost of each work package that they affect (Allahi et al., 2017).

However, this technique has been widely criticised in the literature because this proposed contingency allocation is subjective and based on intuition or experience, and is not rational and, therefore, lacks sound justification (Idrus et al., 2011; Mak and Picken, 2000). Lam and Siwingwa (2017) conclude that the allocation of cost contingencies by a given percentage is a method that is difficult to defend or justify for being based on observation and experience, which makes objectivity non-existent, whereas a statistical regression analysis better suits models with virtually or absolutely no information, as in early project planning stages.

As a result, new methods of different natures emerge to address the problem of cost allocation as contingencies in projects. Baccarini (2005) and Baccarini and Love (2014), conducted their study on the commonest estimation methods. More recently, Islam et al. (2021) presented a comparison of different contingency cost methods/models/tools by detailing the characteristics, advantages and disadvantages of each followed method.

Approaches like parametric regression, probabilistic distribution and simulation, and methods based on artificial intelligence, are used to quantify cost contingencies more objectively and accurately in projects.

Thus, Diab et al. (2017) apply multiple regression to analyse the dependency relationship between predetermined cost contingency amounts and the perceived ratings of risk drivers assessed by project professionals, and Thal et al. (2010) apply multiple regression to predict cost overruns based upon empirical data available prior to contract award for a construction project. This regression analysis type provides a deterministic model and requires data from a considerable number of similar previous projects, which is not always possible for complex infrastructure projects. Other authors, such as Afzal et al. (2020), Idrus et al. (2011), Salah and Moselhi (2015), and Jung et al. (2015), use fuzzy techniques to convert semantic expert opinions into probabilities and percentage cost overruns. Other works focus on probabilistic models (Hoseini et al., 2020; Touran, 2003) or Artificial Neuronal Networks (ANN) (Lhee et al., 2012).

Undoubtedly, one of the most widely used methods is MCS, which has many advantages compared to previous methods. For example, the statistics of the simulated project are obtained computationally without having to operate with analytical functions, and no simplifications in the analysed model are required to generate the system output. As a result, MCS is widely recognised as a valid technique, which means that its results are more likely to be accepted (Vose, 2008). In addition, MCS allows the analysis of the inputs that most affect the system output, which make it the commonest and most applicable tool for quantifying risks in large engineering projects (Liu et al., 2017). By focusing on the estimation of cost contingency reserves, Barraza and Bueno (2007) assign contingencies individually to each work package of the work breakdown structure (WBS). This method facilitates the improvement, efficiency and speed of taking corrective actions. In turn, the contingency management process is more dynamic, provided we avoid having the bulk of our contingency assigned to the total project. In light of this, Hammad et al. (2016) propose a top-down approach to contingency allocation. This approach involves calculating the overall contingencies needed for the project and then allocating a portion to each WBS package. In its study, contingencies are assigned to each work package based on the uncertainty they bring to the project as a whole. Contingency allocation is based on the proportion of the cost of each activity to the total budget. It also takes into account whether activities are on the critical path. The same authors (Hammad et al., 2015) present the results obtained from implementing their methodology to assign contingencies to a project. Other authors propose developing a probabilistic model that incorporates cost uncertainty to allocate contingencies according to the selected confidence level (Touran, 2003).

Barraza and Bueno (2007) calculate cost contingencies using the difference between the maximum percentile of each activity, obtained by MCS, and its expected cost. In any case, this confidence percentile value, which is chosen after applying MCS, usually depends on risk appetite and the level of organisational maturity. Percentiles P50, P70 and P80 are the most frequently used (Eldosouky et al., 2014), while the P80 percentile is the most widespread (Kwon and Kang, 2019; Lorance and Wendling, 2001; Traynor and Mahmoodian, 2019).

Shahtaheri et al. (2017) follow an integrated approach that includes a probabilistic approach to project risks in conjunction with MCS in complex projects. Similarly, Chang and Ko (2017) apply an integrated approach to risk assessment and cost prediction using MCS. They use expert judgements to assess the risks associated with the activity cost in parametric forecasting and estimate the net present value (NPV) by considering the activity cost, revenue and the impact of associated risks. Finally, Maronati and Petrovic (2019) assess the project's uncertainties and risks by taking into account both correlated and uncorrelated variables for cost prediction.

In this paper, we use MCS as a tool to estimate project cost contingencies. To do so, as input variables we employ all the project risks identified in the planning phase (aleatoric, stochastic and epistemic uncertainties). As a result of the simulation, we obtain statistical data about the distribution function of the total project cost. From this point, Project Managers determine the margin of contingencies according to their risk aversion. The cost associated with the selected percentile is compared to the project's planned cost. This method enables us to determine the necessary cost contingencies for the project depending on the uncertainty type associated with each identified risk.

## 3. Research method

To estimate the cost contingencies, we carry out an integrated analysis of programming and cost risk. To do so, we follow the model shown in Fig. 1, which can be divided into three distinct blocks. In the first one (Project Analysis), we use the detailed project schedule information, which includes all the work to be performed, where the precedence relation of each activity is reflected and allows us to build an AON (Activity On Node) graph-type diagram on which we represent the precedence relations of project activities following the recommended CPM (Critical Path Method) scheduling practice (AACE - American Association of Cost Engineering, 2011), which allows calculating the critical paths accurately as a basis for Monte Carlo simulation. In this phase, we plan the estimate of the cost and duration of activities.

The second block of our model corresponds to project risk identification and the subsequent risk assessments. A Risk Register is created that includes all the uncertainties that may impact both project duration and cost. To estimate cost contingencies, we include not only the identified risks that exclusively affect the project's cost but also those risks that impact the duration of the activities making up the project (or impact the project as a whole). The reason for this decision is that longer or shorter activity duration varies the activity's cost and, therefore, the total project cost. When modelling the cost of each project activity, we assume that there are fixed costs (independently of activity duration) and variable costs (proportional to activity duration).

During the risk identification process, we consider the inherent aleatoric uncertainty in project activities. This uncertainty type means that duration and, therefore, the cost can take a random value, which can be modelled by a distribution function (Normal, Beta, Triangular, or others). The Project Risk Management Experts Committee must also identify those probabilistic risk events or situations that may occur and positively or negatively influence project objectives (duration and cost). This uncertainty type (stochastic uncertainty) is included in the risk register for further analyses and assessments before simulation. In the event of it not being possible to assess risk with absolute certainty (with a fixed percentage of probability), then semantic values can be used to identify and estimate it. This uncertainty type (epistemic) is also included as a risk in the register created for the project.

Finally, each identified risk is associated with the activity which it might impact, and it is finally introduced as input variables into MCS. A comprehensive and adequately assessed Risk Register is essential to obtain accurate, correct, and reliable results. The Risk Register should include the occurrence probability of the risk and the impact it might have on the project's objectives (measured as an impact on duration or cost). If the employed data are not quality data, or the risk identification and assessment process are incomplete, the outcome obtained as a contingency value for the project will not be realistic and will not be used confidently by Project Managers.

The third and last block of our model includes MCS as a technique or tool. This tool is used to perform a quantitative analysis of the total project risk. Different commercial tools allow us to perform quantitative risk analyses (@Risk, Crystal Ball, Primavera, etc.). However, none of these tools offers adequate versatility to implement any particular project, or to incorporate the different identified uncertainty types (stochastic, aleatoric, and epistemic). Therefore, we developed an "*ad hoc*" algorithm in Matlab,[1] which allows us to perform MCS and to meet the above requirements. The output of this integrated risk analysis provides graphical representations and statistical data of the distribution functions of the total project duration and cost. In this way, we can know the probability of the duration and cost objectives being met by incorporating the risks that may affect the project. The analysis of the data obtained with MCS permits us to determine the cost contingency reserves that enable us to meet the risk threshold set by the project's management. These reserves depend on Project Managers' risk aversion. To calculate them, Project Managers choose the percentile of the probability distribution functions (PDF) with which they feel more confident.

[1] The developed code has registered Intellectual Property (765–704822). The code is available to users with prior authorization from the author.

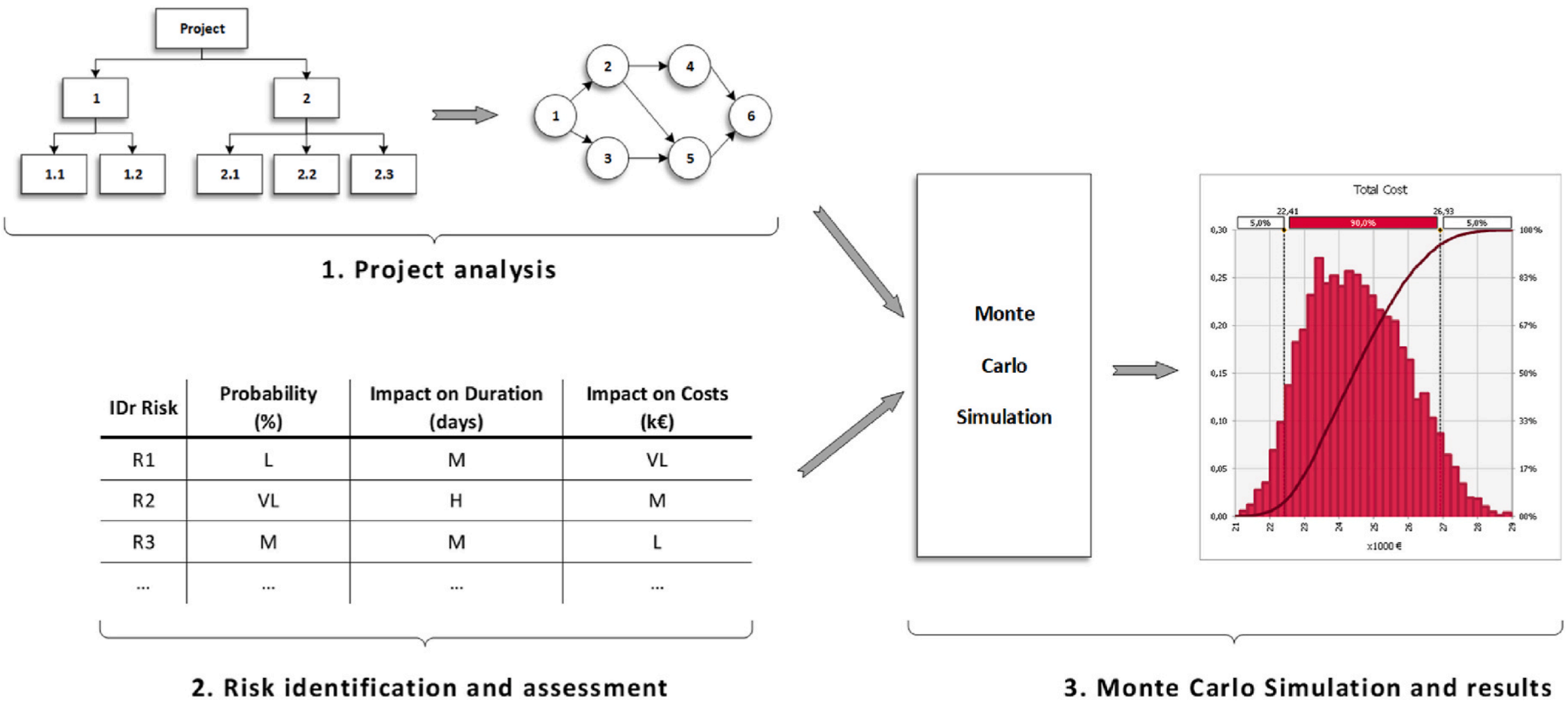

**Fig. 1.** Proposed model for the project cost contingency estimation.

### 3.1. Project analysis

Following the PMBOK guide (Project Management Institute, 2017), we should obtain a project schedule using the project activity list. This list of activities is obtained from the work packages described in the work breakdown structure (WBS) that contains all the work defined to complete the project scope. Based on the list of activities, and in compliance with the constraints imposed by the duration and precedence relation of each activity, we obtain an AON network in which activities are scheduled at their earliest start date based on their precedence relations. These methods provide a schedule in which each project activity is set to start at an earlier time (i.e., activities are scheduled to start at an earlier time).

One key input of the integrated risk analysis is to obtain an accurate realistic estimate of both the duration and cost of activities. The estimated duration and cost values for each activity should be based on the mode or the "most likely" value of each activity, which should be consistent with the schedule and resource availability (Acebes et al., 2021). Notwithstanding, given that risks are identified and assessed by experts, there is a possibility of biases among different experts. As Vasvári (2015) puts it, subjectivity cannot be completely excluded from risk management. This inherent subjectivity affects the willingness to assume risks and, therefore, in the evaluation of risks. However, there are ways to avoid subjective bias. For example, quantitative data can be used whenever possible. If this is not possible, subjective judgements in verbal descriptions can be mitigated by adding quantitative descriptions (e.g., ranges) to the definition of categories (Duijm, 2015).

### 3.2. Risk identification and assessment

The Recommended Practice (AACE - American Association of Cost Engineering, 2011) recommends that a Risk Register is kept in such a way that risk information is updated. The first step is to identify the risk to be assessed. Then it is necessary to include those activities or work packages that are affected by a risk, which may affect the project as a whole. Finally, details of the occurrence probability and its impact on the cost and duration of the set of affected activities should be included.

#### 3.2.1. Risk identification

For the identification process, risk-related information can be collected through workshops and interviews held with risk experts (AACE - American Association of Cost Engineering, 2011; Van et al., 2019). Another way to identify risks, and to complement the information in the register with a breakdown of affected costs, duration, quality, scope and activities or work packages (Seyedhoseini et al., 2009), is by referring to similar projects, lessons learned from other projects in the organisation, engineering estimates, applying expert judgements, market studies, the information provided by suppliers and subcontractors, among others (AACE - American Association of Cost Engineering, 2011; Cagliano et al., 2015).

To assist the risk identification process, we can use different general sources of risks, such as those arising from the studied uncertainty types (Hillson, 2014). In line with this, El-Sayegh (2008) assesses the impact of overall risks on construction projects. Similarly, the study of Sonmez et al. (2007) draws on the idea that the risks inherent to the project in question occur, as do other external environmental risks to the project organisation.

Idrus et al. (2011) divide the risks of a construction project into internal and external risks by adjusting them to each particular case. Although they recognise overall risks, such as currency crises, developer bankruptcy and natural disasters (e.g., earthquakes and tsunamis), and these risks can affect all construction projects, these authors do not include them in their analysis because they are not usually very relevant for the bidding process.

A classification of the overall risks in construction projects proposed by different authors is found in Table 1.

Besides studying overall risks, there are specific risks for the construction sector. Zhi (1995) highlights the need to incorporate a global macro level vision of the political and economic situation in the region where the project is undertaken. At the same time, he recommends including the uncertainty that derives from local and regional particularities in the risk analysis (see Table 2). Han and Diekmann (2001) divide overall risks into five categories: political, economic, cultural/legal, and technological/construction and other risks. Specifically, construction risks arise from entering the international construction market and given differences between the company's place of origin and the location of the construction project.

Bu-Qammaz et al. (2009) make a comparison of different countries to create a category specifically for this purpose. Another important aspect in production processes is the logistical supply chain, which is why Gosling et al. (2013) identify five sources of uncertainty in supply chains.

**Table 1**
Classification of the overall risks associated with a construction project.

| Author | Major Risks | Risk Factors | |
|---|---|---|---|
| El-Sayegh (2008) | Internal Risks | Owners | Contractors |
| | | Designers | Subcontractors |
| | | Suppliers | |
| | External Risks | Political | Economic |
| | | Social & Cultural | Natural |
| | | Others | |
| Sonmez et al. (2007) | Project Risk Factors | General | Contractor |
| | | Contract | Financial |
| | | Design | Site |
| | | Partnership | |
| | Country Risk Factors | Financial | Laws & Regulations |
| | | Political | Administration |
| | | Market Potential | Resources |
| Idrus et al. (2011) | Internal Risks | Safety | Labour Dispute |
| | | Equipment Failure | Unavailability or Resources |
| | | Defective Materials | Mismanagement |
| | | Quality of Work | |
| | External Risks | Different Site Conditions | Change in Government Policy |
| | | Weather Conditions | Changes in Economic Conditions |
| | | Social Impact | Delayed Payment |
| | | Third-Party Delays | |

Whatever the method we use, we obtain a list of risks that can affect the project's time and cost objectives. Risk identification allows us to associate each risk with the activities that it may impact (or even on the project as a whole). Finally, we can assess them.

### 3.2.2. *Risk assessment*

As part of our integrated project risk management process, we need to categorise the probability and severity of the risks identified in the previous step. The probability of risk is defined by a probability distribution functions (PDF), which is responsible for modelling uncertainty. This PDF should reflect the occurrence probability percentage of such an event as realistically as possible (Baccarini and Love, 2014). We can similarly proceed with the impact of a risk. In this case, we differentiate whether the risk impacts the project cost or its duration. Severity is understood as the relevance of uncertainty (Gosling et al., 2013) and is estimated according to the project's objectives that it impacts. During our integrated risk management process, the impact is also modelled (in cost and duration terms) with a PDF, the choice of which depends on the uncertainty type to be represented.

Hillson (2014) distinguishes three uncertainty types that can be modelled and incorporated into MCS: aleatoric, stochastic and epistemic. There is another uncertainty type (ontological), but it lies beyond the limits of human knowledge and, thus, cannot be modelled (Alleman, Coonce and Price, 2018a).

In this paper, we quantify the probability and impact of project risks based on the four uncertainty types proposed by Hillson (2014). They are then used as inputs for the quantitative risk analysis, which we employ to calculate cost contingencies, and they should be set by Project Managers.

*Aleatoric uncertainty* arises from variability (Alleman, Coonce and Price, 2018b; Chapman and Ward, 2004). It is the uncertainty type that is due to the inherent variability of any natural phenomenon owing to multiple causes. The commonest way of modelling aleatoric uncertainty for the duration of activities is through probabilistic distribution functions: Beta, Triangular, Normal and Uniform. Cost variability has been modelled with Beta, Triangular, Lognormal and Pearson-type functions (Ordóñez Arízaga, 2007). An analysis of the probabilistic distribution functions most frequently used in the process to incorporate the variability that derives from aleatoric uncertainty is presented in Table 3.

Event uncertainty, or *stochastic uncertainty,* is based on the occurrence of events with known consequences (Hillson, 2020). These are events for which we know, or can be assured with certain precision, the probability or impact associated with the event occurring. In other words, we have no doubts about the likelihood or impact that it would have on our project, and it would happen if that risk occurring. The most frequent way to deal with this uncertainty type is by studying the most statistically probable future scenarios by positioning simulation as good alternative in the possible contingency estimation (Hillson and Simon, 2012).

**Table 2**
Classification of specific risks for the construction sector.

| Author | Overall Risks | | Construction Risks | |
|---|---|---|---|---|
| Zhi (1995) | Nation/Region | Construction Industry | Market Fluctuations | Contract System |
| | Company | Project | Law and Regulations | Standards and codes |
| Han and Diekmann (2001) | Political Risks | Economic Risks | Different standards | Different measurement system |
| | Cultural/Legal | Other Risks | Material availability | Labour issue (i.e. skill or strike) |
| | Technical/ Construction | Difference in geography | Subcontractor availability | |
| | | | Domestic Requirements | |
| Bu-Qammaz et al. (2009) | Country | Construction | Managerial Complexity | Technical and Technological Complexities |
| | Project Team | Contractual | Subcontractor unavailability | Resources unavailability |
| | Intercountry | Design | Adverse Physical Conditions | Shortage of Client Financial Resources |
| Mahendra et al. (2013) | Technical Risks | Financial Risks | Labour disputes | Design changes |
| | Construction Risks | Socio-Political Risks | Site conditions | Too high-quality standard |
| | Physical Risks | Environmental Risks | Equipment failures | New technology |
| | Organisational Risks | Labour productivity | | |
| Lee et al. (2017) | Political Risks | Other Risks | Construction complexity | PM competency |
| | Economic Risks | Building type | Construction duration | Owner's changes |
| | Social/Cultural Risks | Construction type | Force majeure | Owner's changes |
| | Construction Risks | Contract type for payment | PM competency | |
| Hosny et al. (2018) | External Risks | Subcontractors Risks | Owner-generated Risks | Labour conflicts and disputes |
| | Design Risks | Equipment Risks | Lack of quality management (planning, assurance, control) | Safety issues |
| | Management Risks | Political and Governmental Risks | | Labour cost fluctuation |
| | Construction Risks | | Labour mistakes, rework and idle times | Surveying and site handling mistakes |
| | | Economical Risks | Labour shortage | |

**Table 3**
Probability distribution functions used by several authors to model aleatoric uncertainty.

| Author | Probability Distribution Functions in Monte Carlo Simulation |
|---|---|
| AACE - American Association of Cost Engineering (2011) | Triangular |
| Clark (2001) | |
| Lorance and Wendling (2001) | |
| Para-González et al. (2018) | |
| Mohamed et al. (2020) | Triangular & Uniform |
| Eldosouky et al. (2014) | |
| Traynor and Mahmoodian (2019) | Triangular, Uniform & Lognormal |
| Colin and Vanhoucke (2016) | Lognormal |
| Trietsch et al. (2012) | |
| Acebes et al. (2021); Acebes et al. (2022) | |
| Acebes et al. (2014, 2015, 2020) | Normal |
| Barraza and Bueno (2007) | Normal (Most Likely, 10% Most Likely) |
| Hammad et al. (2016) | Normal (Mean value, 10% Mean value) |

In the same way, as aleatoric uncertainty can be modelled, stochastic uncertainty is also defined by a distribution function. As stochastic uncertainty is associated with risk events (i.e. events with a certain probability of occurring and, if they do occur, they would impact the project), the most appropriate PDF to model this risk type is the Bernoulli distribution because it allows a risk event that may or may not occur to be modelled (Vose, 2008). In line with this, Kwon and Kang (2019) apply MCS to this distribution function type, and they assign a certain probability occurrence percentage of the identified risks.

*Epistemic uncertainty* arises from a lack of knowledge and information about a system or its environment (Alleman et al., 2018b; Damnjanovic and Reinschmidt, 2020). In aleatoric and stochastic uncertainties, probability and impact can be modelled using distribution functions, but as epistemic uncertainty arises from lack of knowledge, we cannot provide an exact value for probability and/or an impact but must use a range of values. Each range specified for probability and impact is project-specific because it depends on the context in which the project is undertaken. Subsequently, each defined interval is associated with a semantic definition (scale). This scale is used to identify the probability and impact of risks. So we can use tables with defined scales to help us to evaluate not only the impact on any of the project objectives (in the most general case), but also duration and cost (in the particular case of our study). The number of levels to be defined on the scale, and the width of the probability and impact intervals, are defined by Project Managers to reflect each organisation's risk appetite (Project Management Institute, 2017). Fig. 2 depicts an example by specifying five proposed levels: 'Very Low', 'Low', 'Medium', 'High' and 'Very High', including the sixth level ('Nil') as a zero risk. With the categorisation provided by this type of tables, we can assess the probability and impact of the risks arising from epistemic uncertainty on the different project dimensions. In the specific case of our study, which seeks to estimate contingency reserves, we focus on the information about duration and cost provided by this type of tables. This information will allow the identification of the range of values that corresponds to the semantic value (scale) that it corresponds to.

As for the idea of modelling the epistemic uncertainty of risks as statistical distribution functions, some authors report three semantic categories, while others use five. Table 4 shows the number of intervals employed by several authors to define probability and impact scales. Table 4 also indicates the assessment method followed to assess risks, as well as the probabilistic distribution functions used to define and model each semantic category. In this paper, we apply uniform distribution functions to model epistemic uncertainty, as proposed by Eldosouky et al. (2014). However, other authors choose the triangular function as a model to represent epistemic uncertainty when they incorporate it into a simulation model (Hulett, 2012).

### 3.3. Monte Carlo Simulation and results

We hold information about the project, which comprises the number of activities, their sequencing and the planned values of duration and cost of each activity (see Section 3.1 Project Analysis). We also have the risks associated with each activity included in the Risk Register. We estimate the occurrence probability of each identified risk, its impact on the project's objectives in duration and cost terms, and if the risk would occur. To this end, PDFs are assigned to represent the uncertainty associated with each risk (see Section 3.2 Risk Identification and Assessment). Finally, we associate each identified risk with its corresponding affected activities and introduce the result as input into our MCS algorithm. Therefore, the final objective is to perform a quantitative analysis based on the MCS. This analysis allows us to subsequently obtain the PDF of the total project cost. We also collect the data corresponding to the different percentiles associated with the PDF of the total project cost.

**Table 4**
Epistemic uncertainty: probability distribution functions (PDF) and number of intervals.

| Author | Number of Intervals | Method | PDF |
|---|---|---|---|
| Han and Diekmann (2001) | 5 | Go/No-Go Decision Model | Triangular |
| El-Sayegh (2008) | 5 | Statistical Correlation | – |
| AACE - American Association of Cost Engineering (2011) | – | MCS | Triangular |
| Idrus et al. (2011) | 5 | Fuzzy Expert System | – |
| Eldosouky et al. (2014) | – | MCS | Uniform |
| Allahi et al. (2017) | 3 | MCS | Triangular |
| EL-Matbaegy et al. (2017) | 5 | Statistical Correlation | – |
| Hosny et al. (2018) | 5 | Probability & Impact Matrix | – |
| Moreno-Cabezali and Fernandez-Crehuet (2020) | 5 | Fuzzy Logic-Based Model | – |

| SCALE | PROBABILITY | +/− IMPACT ON PROJECT OBJECTIVES | | |
|---|---|---|---|---|
| | | TIME | COST | QUALITY |
| Very High | >70% | >6 months | >$5M | Very significant impact on overall functionality |
| High | 51-70% | 3-6 months | $1M-$5M | Significant impact on overall functionality |
| Medium | 31-50% | 1-3 months | $501K-$1M | Some impact in key functional areas |
| Low | 11-30% | 1-4 weeks | $100K-$500K | Minor impact on overall functionality |
| Very Low | 1-10% | 1 week | <$100K | Minor impact on secondary functions |
| Nil | <1% | No change | No change | No change in functionality |

**Fig. 2.** Example of defining scales for probability and impact. Source: The Project Management Institute (Project Management Institute, 2017).

We decided to use the Matlab programming software application, which allows us to perform NCS by including the different analysed uncertainty types (aleatoric, stochastic, and epistemic). In addition, the algorithm enables projects to be simulated with many activities to be dynamically carried out, but without spending too much time programming the project model to be simulated.

MCS produces a graphical representation of the PDF of the total project cost. We can also obtain the statistical data corresponding to this graph. The MATLAB application provides us with the top percentiles of MCS and the cumulative distribution plots for cost duration. The obtained graphs are analysed using the results table for each percentile.

Finally, the results obtained by considering the most likely future scenarios are compared to those planned for the project implementation in which risks would not occur. This step helps to understand the relevance of external risk in projects and how important it is to practice integrated risk management. These most likely future scenarios are given by the estimated cost values for each selected percentile. The main objective is to define an appropriate percentile for cost contingencies and to compare it to the contingencies that would be obtained by applying the traditional method, which consists of reserving a certain percentage of the total planned cost. This conventional method focuses on general projects and ignores any particularities that stem from the project's context. This work aims to include these particularities when selecting a budgetary reserve to address the risks that are generated throughout the project life cycle.

## 4. Case study. The "Aulario IndUVa" construction project

In this section, by applying integrated risk management we intend to show how we can set a margin for contingencies in both cost and schedule terms, which is more appropriate to the reality of such a building. To do so, we consider all the risks analysed by the project management team and consider the different risk aversion levels that the organisation may have. Finally, we compare these results to the initial decision made by the developer.

### *4.1. Project analysis*

The "Aulario IndUVa" is a building at the University of Valladolid (Spain), where classes are held in the School of Industrial Engineering (Fig. 3). While the building was being built, contingency margins were set at 10% of the total project cost.

The project consists of 14 main activities. For simplification reasons, 59 project activities are summarised as 14 main activities (or work packages). Information about final activities is omitted, and the main activities are presented instead of aggregate data. The planned project start date is December 2, 2016. The estimated completion date is June 1, 2018. The planned total project duration is 383 working days, which is the unit of measurement of this project. In Fig. 4 we represent the project's Gantt chart, including project activities and their sequencing throughout the execution period.

The project cost is defined by the sum of the costs of project activities. Table 5 shows the estimated duration for each task. It includes the fixed cost, related to the material cost used in each activity; the variable cost, related to the labour cost and, therefore, depending on activities' duration; the total cost, which is the sum of both.

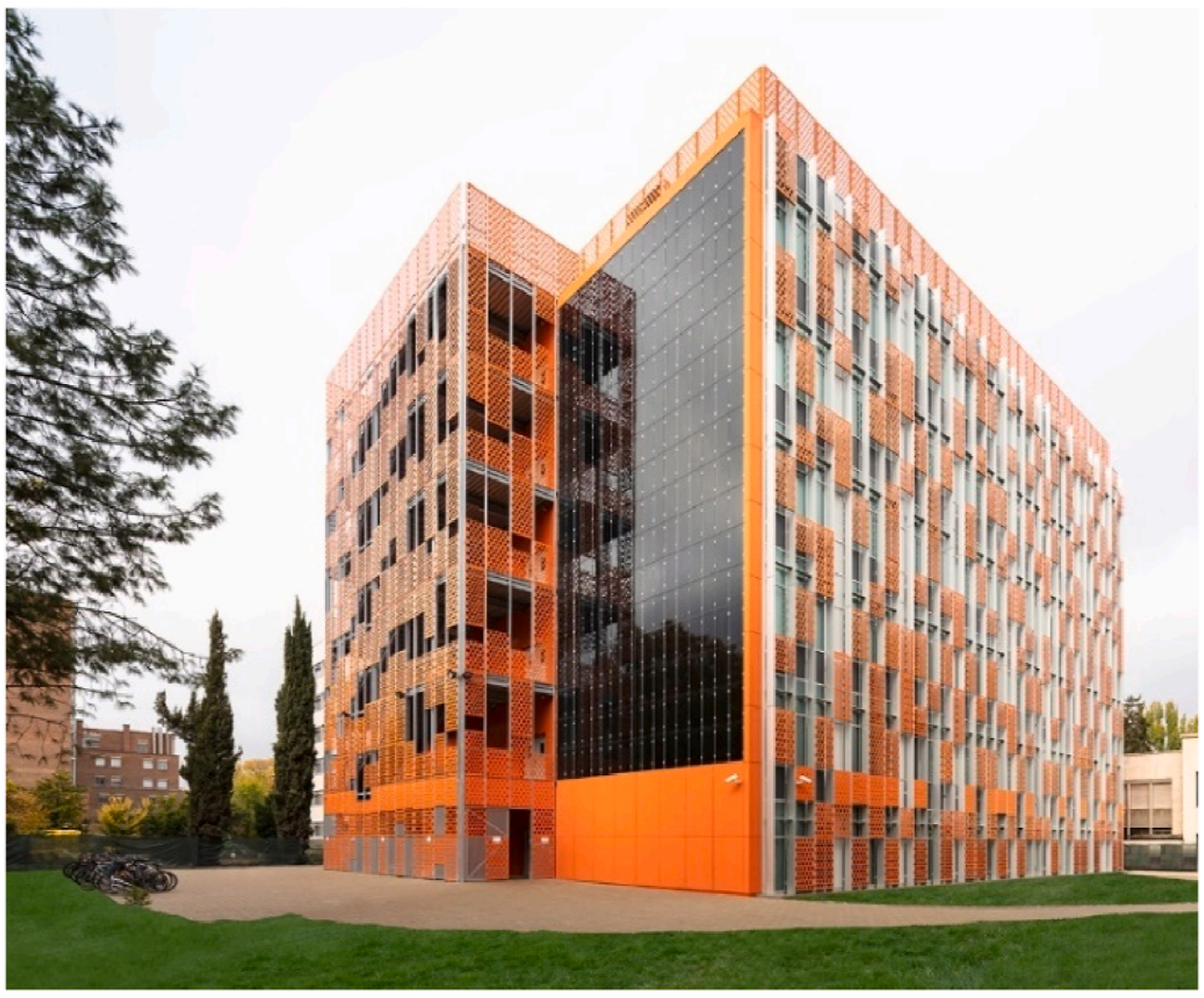

**Fig. 3.** Profile view "Aulario IndUVa".

Table 5 shows that the total planned project cost amounts to €3,935,888.65. The project promoter establishes margins for contingencies of 10% of the total cost, which is equivalent to €393,588.86.

### *4.2. Risk identification and assessment*

#### *4.2.1. Risk identification*

We used different risk identification techniques, such as interviews, experts' judgement of subject matters, risk registers and others. Five people constitute the Experts Committee for this project: the Project Manager, the Construction Manager, and the Construction Supervisors, who are responsible for structures, installations and equipment. As a result, we obtained information on the risks that may impact project objectives. The identified project risks are listed in Table 6. They are related to their particular identification (IDr), and the reason for their possible occurrence is briefly explained (Remarks). Furthermore, the involved risk type is added: threat (T) or opportunity (O), as are the activities that might be affected by the risk if it occurs. Finally, the identifier of the activity (IDact) that may be affected by the risk is included. This code coincides with that shown on the list of activities in Table 5.

Table 6 shows which project activities would be affected by the identified risks associated with stochastic or epistemic uncertainties if they were to occur. It is important to note that there may be activities for which the Experts Committee has not found any risk (associated with stochastic or epistemic uncertainties) that could directly affect these activities. In our case, none of the identified risks directly affects activities 9 (Roofing) or 12 (Urbanisation), which is why these activities do not appear in Table 6.

There is also the possibility of some identified risks not having a direct impact on a specific activity, but on all of them as a whole (i.e., impact on the entire project). This is the case of the risks identified as 8 (Accident) and 11 (Change in the price of raw materials).

To analyse project cost contingencies, not only the risks that may impact the project cost objectives, but also those that may influence the duration of activities, are identified because the latter have an indirect impact on the total project cost.

Beyond the risks associated with stochastic and epistemic uncertainties (i.e., the risks included in Table 6), all the project activities have been considered to have aleatoric uncertainty (i.e., uncertainty due

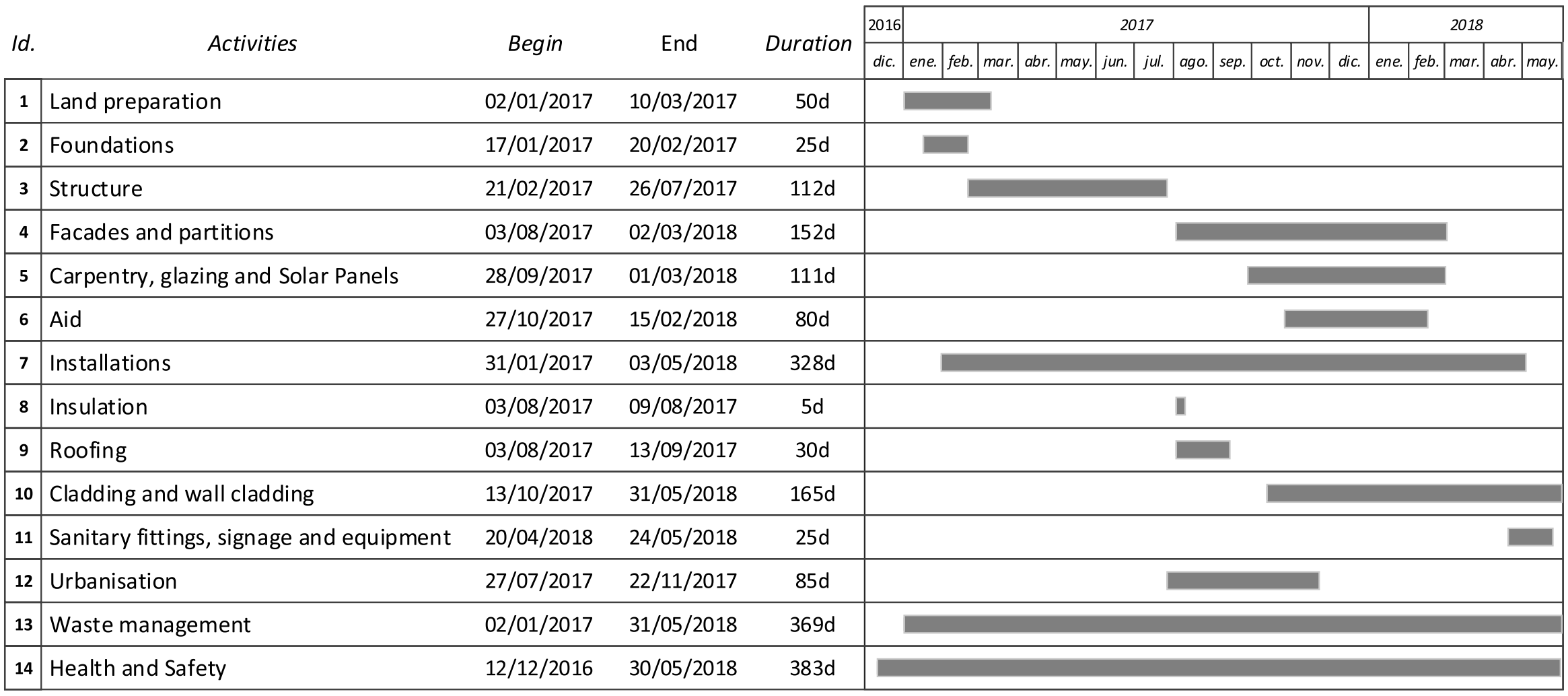

**Fig. 4.** The project's Gantt chart.

**Table 5**
Project duration and cost to build "Aulario IndUVa".

| IDact | Activities | Duration | Variable Cost | Fixed Cost | Total Cost |
|---|---|---|---|---|---|
| 1 | Land preparation | 50 | €52,976.76 | €48,821.90 | €101,798.66 |
| 2 | Foundations | 25 | €10,463.36 | €53,764.17 | €64,227.53 |
| 3 | Structure | 112 | €138,017.78 | €849,450.16 | €987,467.94 |
| 4 | Façades and partitions | 152 | €73,301.94 | €782,100.02 | € 855,401.96 |
| 5 | Carpentry, glazing and solar panels | 111 | €10,211.80 | €194,070.02 | €204,281.82 |
| 6 | Aid | 80 | €30,872.00 | €1,020.00 | €31,892.00 |
| 7 | Installations | 328 | €59,344.20 | €1,010,772.68 | €1,070,116.88 |
| 8 | Insulation | 5 | €12,832.99 | €91,087.44 | €103,920.43 |
| 9 | Roofing | 30 | €14,193.38 | €68,984.83 | €83,178.21 |
| 10 | Cladding and wall cladding | 165 | €48,665.85 | €120.00 | € 353,069.95 |
| 11 | Sanitary fittings, signage and equipment | 25 | €204.84 | €9,005.40 | €9,210.24 |
| 12 | Urbanisation | 85 | €3,932.67 | €13,895.00 | €17,827.67 |
| 13 | Waste management | 369 | €1,900.00 | € 25,783.27 | €27,683.27 |
| 14 | Health and safety | 383 | €4,894.28 | €20,917.81 | € 25,812.09 |
| | *Project* | 383 | *€461,811.84* | *€3,474,076.81* | *€3,935,888.65* |

to the random nature in the duration of activities), which we discuss in the next section.

#### 4.2.2. *Risk assessment*

To carry out this work, we study the different uncertainty types used to model each risk. We first define the criteria that, in turn, define the various uncertainty types and, subsequently, we associate each identified risk with the uncertainty type that can best model its behaviour. The Experts Committee specifies the characteristic parameters of each distribution function assigned to each identified risk for both their probability of, and impact on, duration and/or cost.

We incorporate **aleatoric uncertainty** into the duration of the activities modelled according to a lognormal distribution function (see Table 3), as proposed by Colin and Vanhoucke (2016), Traynor and Mahmoodian (2019) and Trietsch et al. (2012). The lognormal distributions that we use to model activities' aleatoric uncertainty are defined by the estimated duration of each activity and its variance. Variance corresponds to a random number bound within the interval [0.1, 0.3]. This variability range was suggested by Ballesteros-Pérez et al. (2019) after analysing a database of 101 construction projects. This work identified that most of the variability of activity durations fell within the range with a value of 0.10 for activities with low variability, 0.20 for medium variability and 0.3 for activities with high variability. Thus, according to the above-cited authors, the amount of introduced uncertainty comes close to the error made when estimating the duration of tasks in construction projects.

The activity cost comprises a fixed cost and a variable cost, where the latter depends on activity duration. Therefore, the aleatoric uncertainty in the activity cost stems from activity duration variability.

Aleatoric uncertainty can also be used to model some known risks characterised by their variability. In this case, we do not know the exact probability value, but we can define some PDF type to model its behaviour (triangular, normal, beta, other). The occurrence probability of the risk is determined by the distribution function assigned to that activity. In the same way, and independently, it is possible to model the activity's impact with a distribution function (cost impact and schedule impact). If this is the case, this occurs to a greater or lesser extent depending on the model previously defined for that impact.

Risks associated with **stochastic uncertainty** have a known occurrence probability. This occurrence probability will be treated with a Bernoulli distribution function (Kwon and Kang, 2019). According to Allahi et al. (2017), this is the most frequent way to introduce stochastic uncertainty into risk analyses. Analogous (and independent) to the probability modelling, the impact (in duration and cost terms) will be modelled using a distribution function. The distribution function that models the impact will be defined by the project's Experts Committee. As shown in Table 9, the distribution function that models the impact may be triangular, uniform, or even deterministic value if the effect on the activity's duration or cost is a constant value.

To model **epistemic uncertainty**, we firstly define the probability and impact range required for this particular project (Tables 7 and 8 below). Subsequently, we assign each identified risk with an estimated probability and impact range (in cost and/or duration terms). This risk type is characterised by not precisely knowing the occurrence probability or the magnitude of the possible impact. Thanks to the experience gained from previous projects, the project's Experts Committee assesses each risk by identifying it with a qualitative term: 'Very Low', 'Low', 'Medium', 'High' and 'Very High'.

**Table 6**
Risks identified in the project.

| IDr | Risk | Remarks | Type | ActID | Activity affected |
|---|---|---|---|---|---|
| 1 | Labour availability | Subcontractors may have other staffing needs and their involvement in the project can be affected | T/O | 3<br>4<br>5<br>8 | Structure<br>Façades and partitions<br>Carpentry, glazing, and solar panels<br>Insulation |
| 2 | Materials and equipment availability | There may be delays in certain strategic supplies to secure the project, such as restrictions to import goods or blocking international trade. | T | 5<br>7<br>10<br>11 | Carpentry, glazing, and solar panels<br>Installations<br>Cladding and wall cladding<br>Sanitary fittings, signage, and equipment |
| 3 | Legalising facilities | Local authorities may take a long time to legalise facilities due to red tape | T | 10 | Cladding and wall cladding |
| 4 | Archaeological remains | The discovery of archaeological remains can stop work until it is determined whether they are of historic-cultural relevance | T | 1<br>2 | Land preparation<br>Foundations |
| 5 | Water at the phreatic level | The existence of water pockets at the soil phreatic level can stop work until they are eliminated, and work is allowed to continue with foundations. | T | 1<br>2 | Land preparation<br>Foundations |
| 6 | Inclement weather | Severe frost, snowfall and floods can stop the project in its early foundation stages by preventing machinery and operators from working | T/O | 2 | Foundations |
| 7 | Rocks in subsoil | The appearance of rocks in subsoil can stop work until they have been removed | T | 1<br>2 | Land preparation<br>Foundations |
| 8 | Accident | A work accident can immediately stop work for an indefinite time until the reasons for it are clear and the additional safety measures necessary to restart activity are established. | T | | Entire project |
| 9 | Lack of documentation | Lack of the mandatory and necessary documentation to do any construction work can prevent work from being carried out as normal. Such documentation includes registering self-employed and paid workers in the Social Security, or special work permits | T | 5<br>11<br>14 | Carpentry, glazing, and solar panels<br>Sanitary fittings, signage, and equipment<br>Health and safety |
| 10 | Unexpectedly finding asbestos | The unexpected and undocumented appearance of asbestos on plants or asbestos stopping work until it is correctly removed given its severe danger to public health | T | 1 | Land preparation |
| 11 | Change in the price of raw materials | The price of raw materials is subject to constant variations on international markets, which can sometimes mean having to review the prices agreed with subcontractors | T | | Entire project |
| 12 | Changes in regulations | Changes in the legal system can offer an opportunity insofar as it generates the urgency to comply with current legislation and to avoid having to adapt to new proposed changes. | O | 14 | Health and safety |
| 13 | Foundation measurement problems | Measurements of building foundations may not conform to reality because of variations in conditions or the project's environment | T | 2 | Foundations |

These semantic categories have previously been numerically defined by identifying each section with a specific value. Once segments are numerically defined, classes can be modelled by employing distribution functions. We know that, according to estimates, the risk occurrence level lies within a range of probabilities. But since we do not know which of the values within that range best fits the probability of risk occurrence, we employ a uniform probability function to model epistemic uncertainty (i.e., we assume that the probability of occurrence of a risk lies within an equiprobable range of values; Bae et al., 2004). Probabilistic representations of uncertainty have been successfully employed with uniform distributions to characterise uncertainty when knowledge is scarce or does not exist (Helton et al., 2006; Vanhoucke, 2018).

**Table 7**
Probability levels for the analysed project.

| Level | Probability |
|---|---|
| Very Low (VL) | 0–5% |
| Low (L) | 5–12% |
| Medium (M) | 12–20% |
| High (H) | 20–35% |
| Very High (VH) | 35–100% |

**Table 8**
Impact scales in duration and costs terms for the analysed project.

| Level | Impact on Duration | Impact on Cost (€) |
|---|---|---|
| Very Low (VL) | 0–5 days | 0-3k |
| Low (L) | 5–20 days | 3-10k |
| Medium (M) | 20–40 days | 10-25k |
| High (H) | 40–65 days | 25-60k |
| Very High (VH) | >65 days | >60k |

For this particular case study, the project's Experts Committee set the range of values corresponding to each semantic probability value as indicated in Table 7. As discussed above, we assigned a uniform PDF to each semantic level of probability in the risks associated with epistemic uncertainties. A uniform PDF is characterised by two parameters: the minimum possible value and the maximum possible value. At each probability level, the minimum (maximum) value of the uniform PDF corresponding to that level will be the lowest (highest) level within that range.

For example, if the Experts Committee considers that there is a high probability of a risk occurring (level H in Table 7), then we will model the probability of that risk occurring as a uniform random variable with a minimum value of 0.2 and a maximum value of 0.35.

In the same way, it is necessary to define the different intervals (scales) related to the impact on project objectives in terms of both duration (measured in days) and costs (expressed as €). Table 8 shows the definition of impact on this project for epistemic risks.

As each project is unique, with its different scope and objectives, it is important to note that the specific subdivision into levels in probability terms (Table 7), and into impact on costs and durations (Table 8), is defined specifically for each particular project. Consequently, it is possible for the Experts Committee to decide to use different tables in distinct projects to quantify both probability and impact.

After identifying all the risks that affect the project and establishing instructions for assessing all these risks, all this information is integrated into Table 9. With the information on the risks identified in Table 9 and by assigning each risk with its corresponding activity (according to Table 6), we carry out MCS and obtain the simulation results for subsequent analyses.

The above information will be used as input to MCS. To ensure accuracy in the simulation results, it is important that certain basic assumptions are met. The PDFs that model the behaviour of the input variables should come as close as possible to the expected behaviour of the identified risks depending on the type of uncertainty (aleatoric, stochastic or epistemic) that generates those risks. Furthermore, it is assumed that the parameters characterising these random variables (minimum, maximum, average, most likely, etc.) provided by the Experts Committee are accurate. As subjectivity cannot be completely excluded from risk management, it may affect the willingness to assume risks and, therefore, the evaluation of risks. Consequently, it is important to mitigate subjective judgements as discussed above to ensure accurate results.

### 4.3. Monte Carlo simulation and results

To carry out the MCS, we use all the risks identified in the project and listed in Table 9 as the input variables. We also model each project activity with aleatoric uncertainty by assigning a lognormal probability distribution function (PDF). We allocate a distribution function to the occurrence probability of each risk in the Risk Register. We also specify a distribution function to the possible impact on the project cost, and a different one for the impact on project duration. As a result of the simulation, we obtain the principal statistical data of the output distribution functions for the total project cost (Table 10).

**Table 9**
Risk Register of the "Aulario IndUVa" construction project.

| IDr | Risk | Type | NP | PDF | Min | MP | Max | Impact on Duration (days) | | | | | | Impact on Costs (k€) | | | | | |
|---|---|---|---|---|---|---|---|---|---|---|---|---|---|---|---|---|---|---|---|
| | | | | | | | | Type | NRd | PDFd | Min | MP | Max | Type | NRc | PDFc | Min | MP | Max |
| 1 | Labour availability | Aleatoric | | Triang | 85 | 95 | 100 | Aleatoric | | Triang | −5 | 5 | 10 | – | – | – | | – | |
| 2 | Materials and equipment availability | Aleatoric | | Triang | 85 | 95 | 100 | Aleatoric | | Uniform | 0 | | 5 | – | – | – | | – | |
| 3 | Legalising facilities | Epistemic | VH | Unif | 35 | | 100 | Epistemic | VL | Uniform | 20 | | 40 | Epistemic | L | Uniform | 0 | | 3 |
| 4 | Archaeological remains | Epistemic | L | Unif | 5 | | 20 | Aleatoric | | Triang | 30 | 80 | 100 | Aleatoric | | Uniform | 50 | | 100 |
| 5 | Water at the phreatic level | Epistemic | M | Unif | 12 | | 20 | Epistemic | VL | Uniform | 5 | | 20 | Epistemic | M | Uniform | 10 | | 25 |
| 6 | Inclement weather | Aleatoric | | Triang | 85 | 95 | 100 | Aleatoric | | Triang | −5 | 5 | 15 | – | – | – | | – | |
| 7 | Rocks in subsoil | Epistemic | L | Unif | 5 | | 12 | Epistemic | VL | Uniform | 5 | | 20 | Epistemic | H | Uniform | 25 | | 60 |
| 8 | Accident | Epistemic | VL | Unif | 0 | | 5 | Epistemic | VL | Uniform | 0 | | 5 | Epistemic | H | Uniform | 25 | | 60 |
| 9 | Lack of construction documentation | Stochastic | | Bernoulli | | 25 | | Determin | | | | 10 | | Aleatoric | | Uniform | 5 | | 10 |
| 10 | Unexpectedly finding asbestos | Epistemic | VL | Unif | 0 | | 5 | Epistemic | H | Uniform | 40 | | 65 | Epistemic | VH | Uniform | 60 | | 80 |
| 11 | Change in the price of raw materials | Aleatoric | | Triang | 85 | 95 | 100 | – | – | – | | – | | Aleatoric | | Triang | 62 | 77.5 | 93 |
| 12 | Changes in regulations | Epistemic | VL | Unif | 0 | | 5 | Determin. | | | | −10 | | – | – | – | | – | |
| 13 | Foundation measurement problems | Stochastic | | Bernoulli | | 20 | | Aleatoric | | Triang | 7 | 10 | 20 | Aleatoric | | Uniform | 30 | | 60 |

We can see that the average cost of the simulated projects is higher than the project's planned cost. Variance indicates the dispersion of the cost values of simulations concerning the mean value. To carry out a more in-depth study of the data obtained from simulation, we include in Table 11 the data that belongs to the most significant percentiles of the project's cost.

The most striking results indicate that, with a 50% probability (P50), the project costs €4,080,502. This cost represents an increase of €144,614 over the planned value. What is more surprising are the data corresponding to the percentile of the cost planned value. Thus, the percentile to which the budgeted cost (€3,935,888) corresponds is P0.76. Fewer than 1% of the simulated projects (simulation includes the risks identified in the Risk Register) cost less than the budgeted cost. The percentile corresponding to the project's cost, to which we allocate 10% of the budget as the cost contingency margin (3,935,888 + 10% = 4,329,477), results in a percentile of 97.31%. This means that if we allocate a 10% contingency margin to the base budget, 97.31% of the simulated projects cost less than that amount.

Fig. 5 contains the graphical information provided by the application as a result of the MCS. In the same graph, this figure incorporates the probability distribution function together with the cumulative distribution function about the total project cost.

## 5. Discussion

Unlike other quantitative risk analysis methods proposed in the literature that only consider aleatoric uncertainty, in this paper we propose a method that not only considers aleatoric uncertainty, but also the epistemic and stochastic uncertainties associated with each risk identified in the project. In Subsection 5.1, we analyse by means of MCS the effect of incorporating the stochastic and epistemic uncertainties on the project's duration and total cost. In Section 5.2, we employ this information to determine a contingency reserve that better fits the project's reality because it distinguishes the types of uncertainty associated with the risks that affect the project, and it also considers the Project Manager's risk aversion.

### *5.1. Simulation results*

An analysis of the total cost graph (Fig. 5) shows a deviation of the curve towards higher cost values. The histogram that corresponds to the probability distribution function clearly shows the impact of distinguishing the different types of uncertainty associated with the identified risks on the total project cost. Not only does the curve no longer take the traditional "Gaussian bell" shape, but it also shows that a significant number of simulation runs have a strong impact on the project cost. Consequently, the histogram shows a much higher total cost than the most probable value (mode) and, of course, the planned one.

We can also obtain the representation of the final state of each simulated project in a time-cost space (Fig. 6). Each point in the time-cost space represents the end-state situation (time-cost) for every simulation run. This graph incorporates, in turn, the probability distribution functions of time-cost on each corresponding axis (x-axis: project duration; y-axis: project cost).

The above graph evidences the effects of considering the different types of uncertainty that affect the identified project risks. This fact causes the duration and (or) the cost of the simulated project to be longer/higher than that planned and is, therefore, far removed from the "normal" point cloud (that generated only by the aleatoric uncertainty of activities). These simulation runs, whose duration is longer, and cost is higher than expected because they are affected by the impact of some identified risks, lead to a deformed cost distribution function curve. We can also observe that the duration distribution function, located on the abscissa axis, presents greater dispersion, and the curve lengthens towards higher duration values with its most probable value. This representation is due to the positive impact (longer project duration) caused by the risks identified in the project.

Fig. 7 represents the final situation (time/cost) of the simulation runs in two different scenarios: Fig. 7 a) with only the aleatoric uncertainty in project activities; Fig. 7 b), with aleatoric uncertainty and other uncertainty types (i.e., epistemic and stochastic uncertainties) associated with the identified project risks that can impact the project.

We observe that the dispersion of the final scenario in the total project cost for a) (aleatoric uncertainty only) is very narrow compared to the dispersion in duration. This behaviour is because the dispersion of the cost values of the simulated projects is caused by variations in the variable project costs (which, in turn, depend on the duration of activities). In this project, most of the activity costs are made up of fixed costs (material costs, equipment, machinery, etc.). The slight importance of variable costs on the total project results in very little final cost variance. In contrast, in Fig. 7 b) we consider the nature of the uncertainty that affects the project risks (i.e., we incorporate epistemic and stochastic uncertainties). The dispersion of the simulation results is remarkable, especially for project costs. This dispersion is seen not only for those simulation runs with shorter total durations, but also for those with a stronger impact on duration. In the latter, the increased total cost is much higher than expected. This cost increase is the consequence of introducing all the types of uncertainty associated with the duration and cost-identified risks into the simulation.

**Table 10**
Principal statistical data of the output distribution functions of the total project cost.

| Magnitude | Cost (€) |
|---|---|
| *Planned value* | *3,935,888* |
| Mean value simulation | 4,093,200 |
| Variance simulation | 8,960,700,000 |

**Table 11**
Percentiles of the simulated project's cost.

| Percentile | Cost |
|---|---|
| 5.0 | 3,967,100.31 |
| 10.0 | 3,986,083.95 |
| 15.0 | 4,001,911.21 |
| 20.0 | 4,017,208.88 |
| 25.0 | 4,030,954.65 |
| 30.0 | 4,042,922.23 |
| 35.0 | 4,053,471.67 |
| 40.0 | 4,062,953.26 |
| 45.0 | 4,071,841.81 |
| 50.0 | 4,080,502.64 |
| 55.0 | 4,089,128.61 |
| 60.0 | 4,098,090.04 |
| 65.0 | 4,107,650.53 |
| 70.0 | 4,118,122.44 |
| 75.0 | 4,129,774.65 |
| 80.0 | 4,146,100.99 |
| 85.0 | 4,173,065.32 |
| 90.0 | 4,228,488.39 |
| 95.0 | 4,298,427.14 |
| 100.0 | 4,596,002.43 |

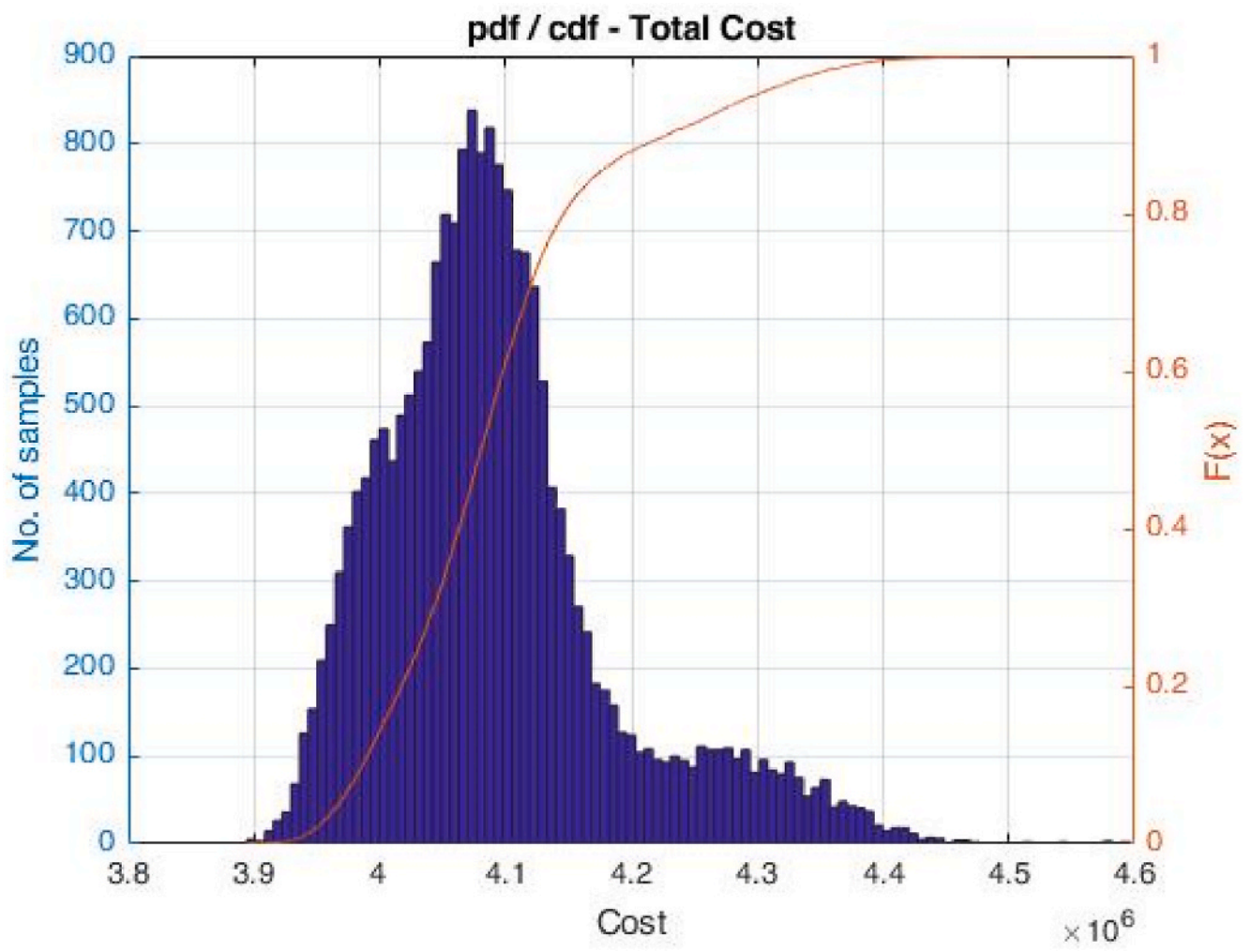

**Fig. 5.** Probability distribution function and cumulative distribution of the total project cost.

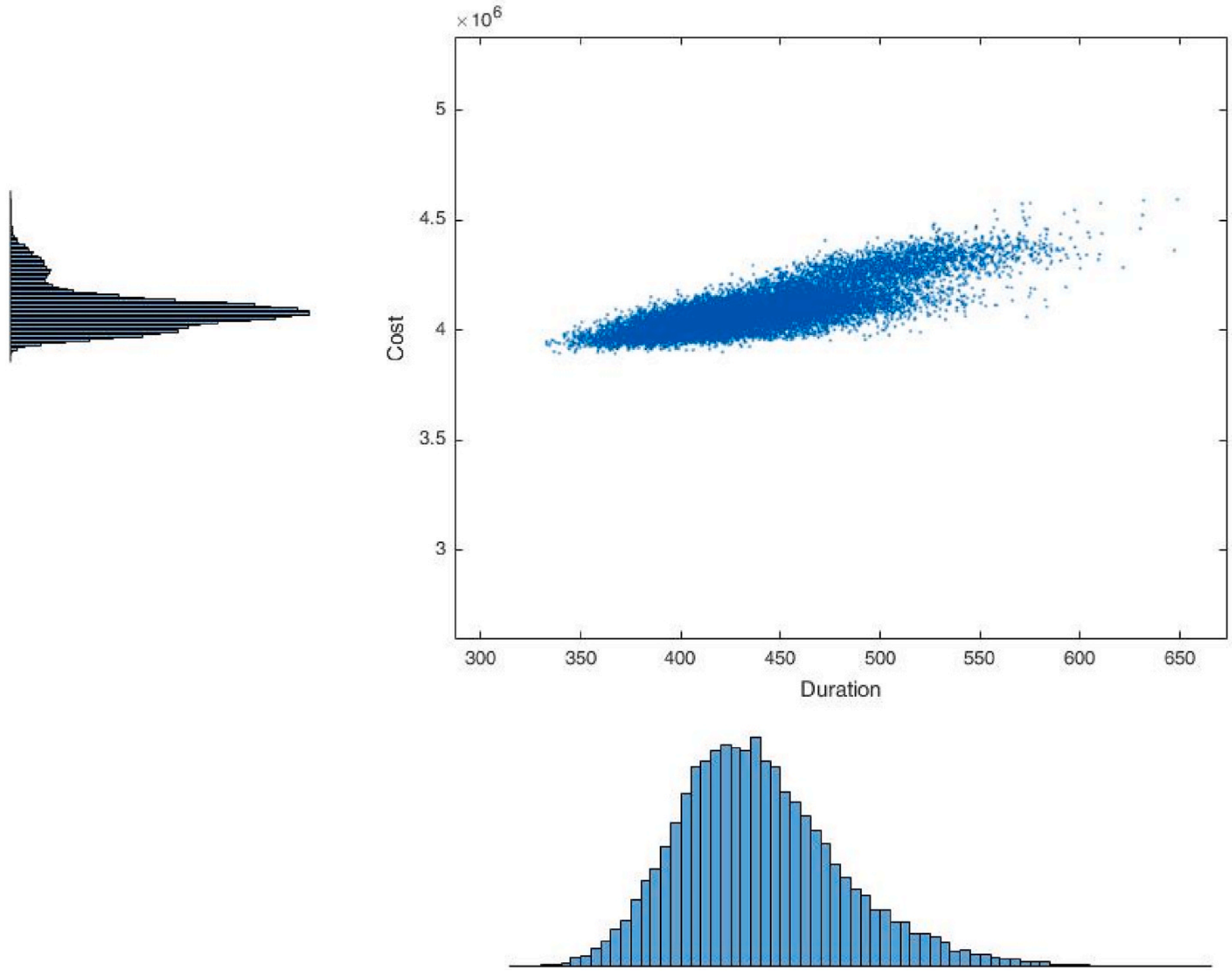

**Fig. 6.** Scatter plot of the simulated projects.

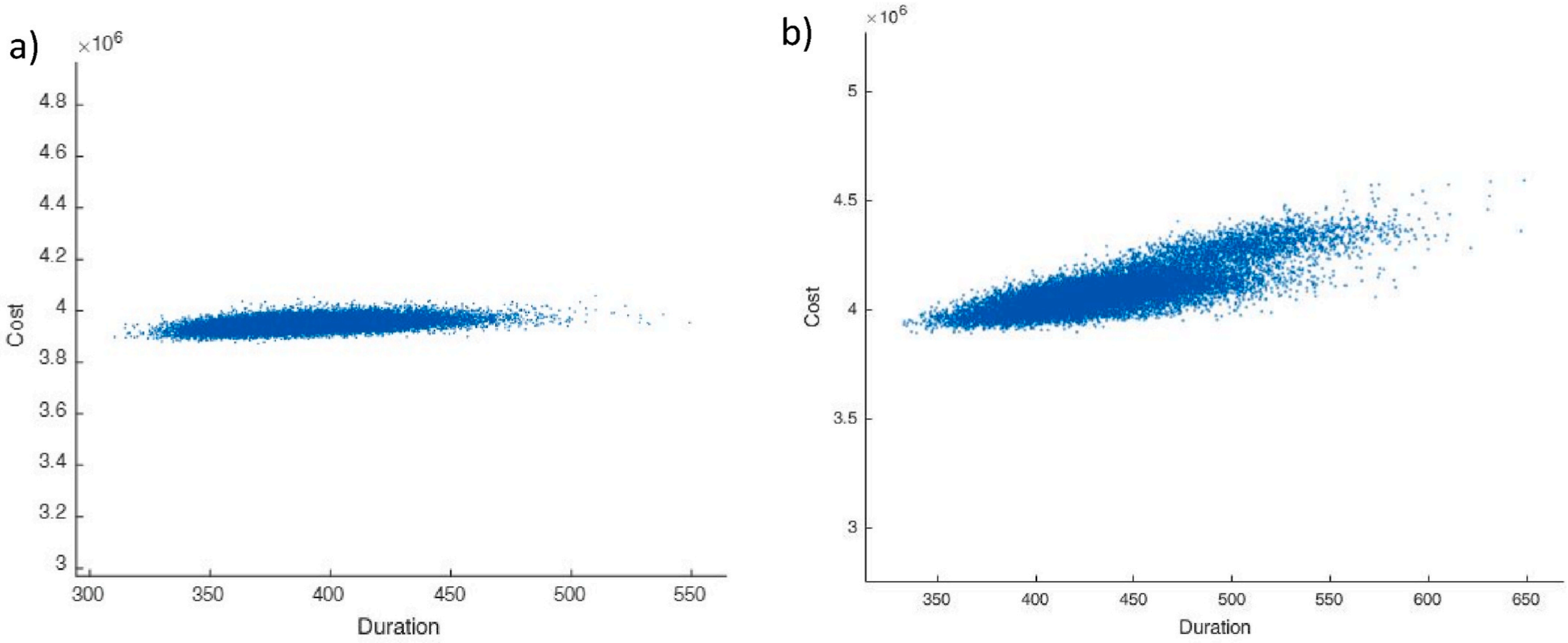

**Fig. 7.** Final scenarios of the simulated projects. a) Aleatoric uncertainty. b) Aleatoric, epistemic and stochastic uncertainties.

Therefore, we demonstrate that it is not sufficient to include only the aleatoric uncertainty of activities to perform quantitative risk analysis based on MCS, as it has traditionally been addressed. Now we see that it is essential to distinguish the different types of uncertainties associated to the risks identified in the project, to adequately model these uncertainties prior to incorporating this information into MCS.

### 5.2. Cost contingency estimation

We now use the MCS results obtained in the previous section (Table 11) to determine whether a cost contingency reserve is appropriate for the risks identified in the project by considering the different types of uncertainty associated with them and the Project Manager's risk aversion. With this information, we allocate a cost contingency reserve calculated as the difference between a percentile according to the Project Manager's risk aversion (Table 11) and the planned project cost (€3,935,888).

In Table 12, we depict several cost contingency reserves based on the simulation results, and also on the Project Manager's risk aversion. We firstly show the value of the planned project cost and the cost contingency estimate used in the construction project under study (i.e., by allocating 10% of the total planned cost as the contingency reserve). Next, we depict several cost percentiles that result from the simulation (taken from Table 11). Then for all these percentiles, we show the cost contingency margin calculated as the difference between the corresponding percentile and the planned cost value of the project.

The rationale behind the proposed method is that, depending on the Project Manager's risk aversion, different reference percentiles can be chosen, which results in distinct contingency reserve values. Thus, the greater the risk aversion, the higher the percentile that should be chosen, and the wider the resulting contingency margin.

For example, if a P70 percentile is chosen (because of a relatively low risk aversion), the result offered by the simulation corresponds to a cost of €4,118,122.44. Consequently, the contingency margin for that percentile will be €182,234.44. If the Project Manager chooses a P80 percentile for the project because of a higher risk aversion, the corresponding project contingency reserve will be €210,212.99 according to the simulation.

Table 12 only includes the P70, P80 and P90 percentiles by way of example of applying the method proposed in this paper. However, the method provides the cost contingency reserve that corresponds to any percentile that we request (i.e., following the Project Manager's risk aversion). Construction projects fall in a very rigid economic sector with a poor risk appetite. Therefore, in most cases, the P80 percentile is selected to estimate the project cost contingencies. In organisations from sectors with more innovation capacity, economic agents tend to have a bigger risk appetite (Baloi and Price, 2003). An organisation's maturity is not the only predictor of risk appetite. There are other attributes that should be considered when assigning a certain maturity level to an organisation (application, processes, organisational risk culture, experience, ability to identify risks, ability to analyse risks, and others) (Zou et al., 2010). However, the business needs to demonstrate a high degree of maturity if a complex risk analysis is to be performed.

The cost contingency reserves obtained by our approach come closer to reality because it considers the costs that arise from simulating those risks modelled according to the actual nature of the (i.e., aleatoric, epistemic or stochastic) uncertainties that generate them. Consequently, these contingency reserves should be more reliable than the estimations made at random or out of "habit" (e.g., a certain percentage of the total planned budget: 10% of the total budget in the construction project under study).

## 6. Conclusions

This paper presents a novel method to allocate cost contingencies in projects. It does so by performing an integrated analysis of the project's cost and duration risks by incorporating all the identified uncertainty types. We should not forget that variability in the duration of activities lead to variability in the costs of these activities and, therefore, in the project. This model can be summarised in three steps: the first step is the project analysis in which the activities that integrate the project are studied, and the project's network diagram is represented. In the second stage, the project risks are identified. All the uncertainties that may impact the cost or duration objectives of any activity should be included in the Risk Register. After a literature review, we classify uncertainties into three types: stochastic, aleatoric and epistemic. We cannot forget another uncertainty type (ontological), whose only response is to reserve a margin of contingencies for these unknown risks.

Having completed the Risk Register for our project, we move on to the last phase of our method. We use the Risk Register and project sequencing data to perform MCS. As a result of the simulation, we obtain the statistics and graphs for total project cost. These results are analysed and, the allocation of project cost contingencies is determined. To make this decision, Project Managers must consider the organisation's degree of risk aversion. The greater the risk aversion, the higher the percentile of the cost function to be chosen to set contingencies.

**Table 12**
Calculating the project's cost contingencies.

| Magnitude | Cost (€) | Contingency Margin (€) |
|---|---|---|
| Planned Project Cost | 3,935,888 | |
| *Actual project "Aulario IndUVa"* | (10%) | 393,588.8 |
| … | … | … |
| P70 | 4,118,122.44 | 182,234.44 |
| P80 | 4,146,100.99 | 210,212.99 |
| P90 | 4,228,488.39 | 292,600.39 |
| … | … | … |

This method was tested on an actual project that consists of a lecture hall construction for the University of Valladolid (Spain). The results obtained with simulation allowed us to compare the allocation of the contingencies that should be reserved if integrated risk analysis is carried out, compared to the traditional method, which consists of allocating 10% of the total planned budget.

The main conclusion drawn from this work is that it is possible to accurately estimate cost contingencies by MCS. This quantitative Risk Management technique enables all the identified sources of uncertainty to be incorporated into the simulation. From the analysis of the obtained results, we conclude that the impact of risks significantly modifies project planning. The difference between the values planned for the project's cost, and the results obtained from carrying out simulation, means that it is necessary to apply the integrated risk management process.

About the comparison made to the traditional method of allocating 10% of the total project budget as cost contingencies, we conclude that the methodology we propose is more specific for quantifying the reserve needed to avoid economically compromising a project's viability. This is because reserves are adjusted to each project's reality and particular context. If simulation provides scenarios with much lower contingencies than that provided by the traditional percentage method, we can allocate the difference to other more profitable investment items. On the contrary, if simulation provides scenarios in which costs are higher than those estimated by the percentage method, we have to set aside more economic resources to ensure the project's feasibility. In other words, we can anticipate future scenarios that are more negative than those planned, which allows us to act accordingly and to increase the probability of the project being a success.

As a future line of research, we think that getting a contingency margin relative to the project schedule is also possible. Following the proposed method, it will be possible to suggest an estimated project completion time after incorporating the identified risks of the project.

## Funding

This research has been partially financed by the Regional Government of Castilla and Leon (Spain) with Grant (VA180P20).

## Data availability

Data will be made available on request.

## Acknowledgements

The authors wish to sincerely thank the Department of Architecture of the University of Valladolid and the company *Constructora San-José S. A.* for providing the necessary information to prepare this document.

## References

AACE - American Association of Cost Engineering, 2011. Integrated cost and schedule risk analysis using Monte Carlo simulation of a CPM model. AACE International Recommended Practice No. 57R-09.

Acebes, F., Pajares, J., Galán, J.M., López-Paredes, A., 2014. A new approach for project control under uncertainty. Going back to the basics. Int. J. Proj. Manag. 32 (3), 423–434. https://doi.org/10.1016/j.ijproman.2013.08.003.

Acebes, F., Pajares, J., González-Varona, J.M., López-Paredes, A., 2020. Project Risk Management from the Bottom-Up: Activity Risk Index. Central European Journal of Operations Research. https://doi.org/10.1007/s10100-020-00703-8.

Acebes, F., Pereda, M., Poza, D., Pajares, J., Galán, J.M., 2015. Stochastic earned value analysis using Monte Carlo simulation and statistical learning techniques. Int. J. Proj. Manag. 33 (7), 1597–1609. https://doi.org/10.1016/j.ijproman.2015.06.012.

Acebes, F., Poza, D., González-Varona, J.M., López-Paredes, A., 2022. Stochastic earned duration analysis for project schedule management. Engineering 9 (February 2022), 148–161. https://doi.org/10.1016/j.eng.2021.07.019.

Acebes, F., Poza, D., González-Varona, J.M., Pajares, J., López-Paredes, A., 2021. On the project risk baseline : integrating aleatory uncertainty into project scheduling. Comput. Ind. Eng. 160 (2021), 107537 https://doi.org/10.1016/j.cie.2021.107537.

Afzal, F., Yunfei, S., Junaid, D., Hanif, M.S., 2020. Cost-risk contingency framework for managing cost overrun in metropolitan projects: using fuzzy-AHP and simulation. Int. J. Manag. Proj. Bus. 13 (5), 1121–1139. https://doi.org/10.1108/IJMPB-07-2019-0175.

Ahmad, I., 1992. Contingency allocation: a computer-aided approach. AACE Transactions, F5 1–7.

Akintoye, A.S., MacLeod, M.J., 1997. Risk analysis and management in construction. Int. J. Proj. Manag. 15 (1), 31–38. https://doi.org/10.1016/S0263-7863(96)00035-X.

Allahi, F., Cassettari, L., Mosca, M., 2017. Stochastic Risk Analysis and Cost Contingency Allocation Approach for Construction Projects Applying Monte Carlo Simulation. World Congress on Engineering. WCE 2017, London. July.

Alleman, G.B., Coonce, T.J., Price, R.A., 2018a. Increasing the probability of program succes with continuous risk management. Coll. Perf. Manag. Meas. News (4), 27–46.

Alleman, G.B., Coonce, T.J., Price, R.A., 2018b. What is Risk? Meas. News 1 (1), 25–34.

Association for Project Management, 2010. Project Risk Analysis and Management Guide (PRAM Guide), second ed. APM Group Limited, Buckinghamshire.

Baccarini, D., 1996. The concept of project complexity - a review. Int. J. Proj. Manag. 14, 201–204. https://doi.org/10.1016/0263-7863(95)00093-3.

Baccarini, D., 2004. Estimating project cost contingency - a model and exploration of research questions. 20th Ann. ARCOM Conf. 1 (September), 105–113.

Baccarini, D., 2005. Estimating Project Cost Contingency-Beyond the 10% Syndrome. Australian Institute of Project Management Conference, Victoria (Australian Institute of Project Management).

Baccarini, D., Love, P.E.D., 2014. Statistical characteristics of cost contingency in water infrastructure projects. J. Construct. Eng. Manag. 140 (3), 04013063 https://doi.org/10.1061/(asce)co.1943-7862.0000820.

Bae, H.R., Grandhi, R.V., Canfield, R.A., 2004. Epistemic uncertainty quantification techniques including evidence theory for large-scale structures. Comput. Struct. 82 (13), 1101–1112. https://doi.org/10.1016/j.compstruc.2004.03.014.

Ballesteros-Pérez, P., Cerezo-Narváez, A., Otero-Mateo, M., Pastor-Fernández, A., Vanhoucke, M., 2019. Performance comparison of activity sensitivity metrics in schedule risk analysis. Autom. ConStruct. 106 (February), 102906 https://doi.org/10.1016/j.autcon.2019.102906.

Baloi, D., Price, A.D.F., 2003. Modelling global risk factors affecting construction cost performance. Int. J. Proj. Manag. 21, 261–269. https://doi.org/10.1016/S0263-7863(02)00017-0.

Barraza, G.A., Bueno, R.A., 2007. Cost contingency management. J. Manag. Eng. (July), 140–146. https://doi.org/10.1061/(ASCE)0742-597X (2007)23:3(140).

Bu-Qammaz, A.S., Dikmen, I., Birgonul, M.T., 2009. Risk assessment of international construction projects using the analytic network process. Can. J. Civ. Eng. 36 (7), 1170–1181. https://doi.org/10.1139/L09-061.

Cagliano, A.C., Grimaldi, S., Rafele, C., 2015. Choosing project risk management techniques. A theoretical framework. J. Risk Res. 18 (2), 232–248. https://doi.org/10.1080/13669877.2014.896398.

Chang, C.-Y., Ko, J.-W., 2017. New approach to estimating the standard deviations of lognormal cost variables in the Monte Carlo analysis of construction risks. J. Construct. Eng. Manag. 143 (1), 06016006 https://doi.org/10.1061/(asce)co.1943-7862.0001207.

Chapman, C.B., Ward, S., 2004. Why risk efficiency is a key aspect of best practice projects. Int. J. Proj. Manag. 22 (8), 619–632. https://doi.org/10.1016/j.ijproman.2004.05.001.

Chapman, C.B., Ward, S., 2011. How to Manage Project Opportunity and Risk. Wiley, New York.

Clark, D.E., 2001. Monte Carlo analysis: ten years of experience. Cost Eng. 43 (6), 40–45.

Colin, J., Vanhoucke, M., 2016. Empirical perspective on activity durations for project-management simulation studies. J. Construct. Eng. Manag. 142 (1), 04015047 https://doi.org/10.1061/(asce)co.1943-7862.0001022.

Damnjanovic, I., Reinschmidt, K., 2020. Data Analytics for Engineering and Construction Project Risk Management. Springer International Publishing, Cham.

Diab, M.F., Varma, A., Panthi, K., 2017. Modeling the construction risk ratings to estimate the contingency in highway projects. J. Construct. Eng. Manag. 143 (8), 04017041 https://doi.org/10.1061/(asce)co.1943-7862.0001334.

Duijm, N.J., 2015. Recommendations on the use and design of risk matrices. Saf. Sci. 76, 21–31. https://doi.org/10.1016/j.ssci.2015.02.014.

El-Kholy, A.M., Tahwia, A.M., Elsayed, M.M., 2020. Prediction of simulated cost contingency for steel reinforcement in building projects: ANN versus regression-based models. Int. J. Construct. Manag. 1–15. https://doi.org/10.1080/15623599.2020.1741492, 0(0).

El-Matbaegy, S., Khalil, M., Sharaf, T., Elghandour, M., 2017. Risk analysis of construction sector in Egypt (during the economic recession periods). Port-Said Eng. Res. J. 21 (2), 37–50. https://doi.org/10.21608/pserj.2017.33231.

El-Sayegh, S.M., 2008. Risk assessment and allocation in the UAE construction industry. Int. J. Proj. Manag. 26 (4), 431–438. https://doi.org/10.1016/j.ijproman.2007.07.004.

Eldosouky, I.A., Ibrahim, A.H., Mohammed, H.E.D., 2014. Management of construction cost contingency covering upside and downside risks. Alex. Eng. J. 53 (4), 863–881. https://doi.org/10.1016/j.aej.2014.09.008.

Elms, D.G., 2004. Structural safety: issues and progress. Prog. Struct. Eng. Mater. 6, 116–126 https://doi.org/https://doi.org/10.1002/pse.176.

Fan, M., Lin, N.P., Sheu, C., 2008. Choosing a project risk-handling strategy: an analytical model. Int. J. Prod. Econ. 112 (2), 700–713. https://doi.org/10.1016/j.ijpe.2007.06.006.

Frank, M., 1999. Treatment of uncertainties in space nuclear risk assessment with examples from Cassini mission implications. Reliab. Eng. Syst. Saf. 66, 203–221. https://doi.org/10.1016/S0951-8320(99)00002-2.

Gosling, J., Naim, M., Towill, D., 2013. Identifying and categorizing the sources of uncertainty in construction supply chains. J. Construct. Eng. Manag. 139 (1), 102–110. https://doi.org/10.1061/(asce)co.1943-7862.0000574.

Hammad, M.W., Abbasi, A., Ryan, M.J., 2015. A new method of cost contingency management. IEEE Int. Conf. Ind. Eng. Eng. Manag. 38–42. https://doi.org/10.1109/IEEM.2015.7385604, 2016-Janua.

Hammad, M.W., Abbasi, A., Ryan, M.J., 2016. Allocation and management of cost contingency in projects. J. Manag. Eng. 32 (6), 1–11. https://doi.org/10.1061/(ASCE)ME.1943-5479.0000447.

Han, S., Diekmann, J., 2001. Approaches for making risk-based go/no-go decision for international projects. J. Construct. Eng. Manag. 127 (4), 300–308. https://doi.org/10.1061/(ASCE)0733-9364 (2001)127:4(300).

Helton, J.C., Johnson, J.D., Oberkampf, W.L., Sallaberry, C.J., 2006. Sensitivity analysis in conjunction with evidence theory representations of epistemic uncertainty. Reliab. Eng. Syst. Saf. 91 (10–11), 1414–1434. https://doi.org/10.1016/j.ress.2005.11.055.

Hillson, D., 2005. In: Describing Probability : the Limitations of Natural Language Dimensions of Risk. PMI® Global Congress 2005—EMEA, vol. 1. Project Management Institute Inc, Edinburgh, Scotland.

Hillson, D., 2014. How to manage the risks you didn't know you were taking. PMI® Glob. Congr. 1–8.

Hillson, D., 2020. Capturing upside risk: finding and managing opportunities in projects. In: Capturing Upside Risk, first ed. Taylor & Francis, Boca Raton.

Hillson, D., Simon, P., 2012. Practical Project Risk Management: the ATOM Methodology, second ed. Management Concepts Inc, Virginia. Tysons Corner.

Hong, J., Shen, G.Q., Peng, Y., Feng, Y., Mao, C., 2016. Uncertainty analysis for measuring greenhouse gas emissions in the building construction phase: a case study in China. J. Clean. Prod. 129, 183–195. https://doi.org/10.1016/j.jclepro.2016.04.085.

Hoseini, E., Bosch-Rekveldt, M., Hertogh, M., 2020. Cost contingency and cost evolvement of construction projects in the preconstruction phase. J. Construct. Eng. Manag. 146 (6), 05020006 https://doi.org/10.1061/(ASCE)CO.1943-7862.0001842.

Hosny, H.E., Ibrahim, A.H., Fraig, R.F., 2018. Risk management framework for Continuous Flight Auger piles construction in Egypt. Alex. Eng. J. 57 (4), 2667–2677. https://doi.org/10.1016/j.aej.2017.10.003.

Hulett, D.T., 2012. Acumen Risk for Schedule Risk Analysis - A User's Perspective. Retrieved May 23, 2021, from White Paper website:

Idrus, A., Nuruddin, M.F., Rohman, M.A., 2011. Development of project cost contingency estimation model using risk analysis and fuzzy expert system. Expert Syst. Appl. 38 (3), 1501–1508. https://doi.org/10.1016/j.eswa.2010.07.061.

Islam, M.S., Nepal, M.P., Skitmore, M., Drogemuller, R., 2021. Risk induced contingency cost modeling for power plant projects. Autom. ConStruct. 123 (February 2020), 103519 https://doi.org/10.1016/j.autcon.2020.103519.

Jung, J.H., Kim, D.Y., Lee, H.K., 2015. The computer-based contingency estimation through analysis cost overrun risk of public construction project. KSCE J. Civ. Eng. 20 (4), 1119–1130. https://doi.org/10.1007/s12205-015-0184-8.

Kwon, H., Kang, C.W., 2019. Improving project budget estimation accuracy and precision by analyzing reserves for both identified and unidentified risks. Proj. Manag. J. 50 (1), 86–100. https://doi.org/10.1177/8756972818810963.

Lam, T.Y.M., Siwingwa, N., 2017. Risk management and contingency sum of construction projects. J. Finance Manag. Prop. Construct. 22 (3), 237–251. https://doi.org/10.1108/JFMPC-10-2016-0047.

Lee, K., Lee, H.-S., Park, M., Kim, D.Y., Jung, M., 2017. Management-reserve estimation for international construction projects based on risk-informed k-NN. J. Manag. Eng. 33 (4) https://doi.org/10.1061/(ASCE)ME.1943-5479.0000510.

Lhee, S.C., Issa, R.R.A., Flood, I., 2012. Prediction of financial contingency for asphalt resurfacing projects using artificial neural networks. J. Construct. Eng. Manag. 138 (1), 22–30. https://doi.org/10.1061/(asce)co.1943-7862.0000408.

Liu, J., Jin, F., Xie, Q., Skitmore, M., 2017. Improving risk assessment in financial feasibility of international engineering projects: a risk driver perspective. Int. J. Proj. Manag. 35 (2), 204–211. https://doi.org/10.1016/j.ijproman.2016.11.004.

Long, L.D., Ohsato, A., 2008. Fuzzy critical chain method for project scheduling under resource constraints and uncertainty. Int. J. Proj. Manag. 26 (6), 688–698. https://doi.org/10.1016/j.ijproman.2007.09.012.

Lorance, R.B., Wendling, R.V., 2001. Basic techniques for analyzing and presentation of cost risk analysis. Cost Eng. 43 (6), 25–31.

Mahendra, P.A., Pitroda, J.R., Bhavsar, J.J., 2013. A study of risk management techniques for construction projects in developing countries. Int. J. Innovative Technol. Explor. Eng. 3 (5), 139–142.
Mak, S., Picken, D., 2000. Using risk analisys to determine construction project contingencies. J. Construct. Eng. Manag. 126 (April), 130–136. https://doi.org/10.1061/(ASCE)0733-9364 (2000)126:2(130).
Maronati, G., Petrovic, B., 2019. Estimating cost uncertainties in nuclear power plant construction through Monte Carlo sampled correlated random variables. Prog. Nucl. Energy 111 (August 2018), 211–222. https://doi.org/10.1016/j.pnucene.2018.11.011.
Mohamed, E., Seresht, N.G., Hague, S., Abourizk, S.M., 2020. Simulation-based approach for risk assessment in onshore wind farm construction projects. *2020 Asia-Pacific International Symposium on Advanced Reliability and Maintenance Modeling.* APARM 2020. https://doi.org/10.1109/APARM49247.2020.9209516.
Moreno-Cabezali, B.M., Fernandez-Crehuet, J.M., 2020. Application of a fuzzy-logic based model for risk assessment in additive manufacturing R&D projects. Comput. Ind. Eng. 145 (May) https://doi.org/10.1016/j.cie.2020.106529.
Moret, Y., Einstein, H.H., 2016. Construction cost and duration uncertainty model: application to high-speed rail line project. J. Construct. Eng. Manag. 142 (10), 05016010 https://doi.org/10.1061/(asce)co.1943-7862.0001161.
Moselhi, O., 1997. Risk assessment and contingency estimating. AACE Int. Trans. 90, 1–6.
Ordóñez Arízaga, J.F., 2007. A Methodology for Project Risk Analysis Using Bayesian Belief Networks within a Monte Carlo Simulation Environment. University of Maryland, College Park.
Para-González, L., Mascaraque-Ramírez, C., Madrid, A.E., 2018. Obtaining the budget contingency reserve through the Monte Carlo method: study of a ferry construction project. Brodogradnja 69 (3), 79–95. https://doi.org/10.21278/brod69305.
Project Management Institute, 2017. A Guide to the Project Management Body of Knowledge: PMBoK(R) Guide, sixth ed. Project Management Institute Inc, Pennsylvania - USA. sixth ed.
Salah, A., Moselhi, O., 2015. Contingency modelling for construction projects using fuzzy-set theory. Eng. Construct. Architect. Manag. 22 (2), 214–241. https://doi.org/10.1108/ECAM-03-2014-0039.
Seyedhoseini, S.M., Noori, S., Hatefi, M.A., 2009. An integrated methodology for assessment and selection of the project risk response actions. Risk Anal. 29 (5), 752–763. https://doi.org/10.1111/j.1539-6924.2008.01187.x.
Shahtaheri, M., Haas, C.T., Rashedi, R., 2017. Applying very large scale integration reliability theory for understanding the impacts of type II risks on megaprojects. J. Manag. Eng. 33 (4), 04017003 https://doi.org/10.1061/(asce)me.1943-5479.0000504.
Sonmez, R., Ergin, A., Birgonul, M.T., 2007. Quantitative methodology for determination of cost. J. Manag. Eng. 23 (January), 35–39. https://doi.org/10.1061/(ASCE)0742-597X, 2007)23:1(35.
Thal, A.E., Cook, J.J., White, E.D., 2010. Estimation of cost contingency for air force construction projects. J. Construct. Eng. Manag. 136 (11), 1181–1188. https://doi.org/10.1061/(asce)co.1943-7862.0000227.
Touran, A., 2003. Probabilistic model for cost contingency. J. Construct. Eng. Manag. 129 (3), 280–284. https://doi.org/10.1061/(asce)0733-9364 (2003)129:3(280).
Traynor, B.A., Mahmoodian, M., 2019. Time and cost contingency management using Monte Carlo simulation. Aust. J. Civ. Eng. 17 (1), 11–18. https://doi.org/10.1080/14488353.2019.1606499.
Trietsch, D., Mazmanyan, L., Gevorgyan, L., Baker, K.R., 2012. Modeling activity times by the Parkinson distribution with a lognormal core: theory and validation. Eur. J. Oper. Res. 216 (2), 386–396. https://doi.org/10.1016/j.ejor.2011.07.054.
Uzzafer, M., 2013. A contingency estimation model for software projects. Int. J. Proj. Manag. 31 (7), 981–993. https://doi.org/10.1016/j.ijproman.2012.12.002.
Van, S.Q., Le-Hoai, L., Dang, C.N., 2019. Predicting implementation cost contingencies for residential construction projects in flood-prone areas. Int. J. Manag. Proj. Bus. 12 (4), 1097–1119. https://doi.org/10.1108/IJMPB-04-2018-0071.
Vanhoucke, M., 2018. The data-driven project manager: a statistical battle against project obstacles. In: The Data-Driven Project Manager: A Statistical Battle against Project Obstacles. https://doi.org/10.1007/978-1-4842-3498-3.
Vasvári, T., 2015. Risk, risk perception, risk management - a review of the literature. Publ. Finance Quart. 60 (1), 29–48, 0031-496X.
Vose, D., 2008. Risk Analysis: a Quantitative Guide, third ed. Wiley, Chichester, U.K.
Zhang, Y., Fan, Z.P., 2014. An optimization method for selecting project risk response strategies. Int. J. Proj. Manag. 32 (3), 412–422. https://doi.org/10.1016/j.ijproman.2013.06.006.
Zhi, H., 1995. Risk management for overseas construction projects. Int. J. Proj. Manag. 13 (4), 231–237. https://doi.org/10.1016/0263-7863(95)00015-I.
Zou, P.X.W., Chen, Y., Chan, T., 2010. Understanding and improving your risk management capability : assessment model for construction organizations. J. Construct. Eng. Manag. 136 (August), 854–863. https://doi.org/10.1061/(ASCE)CO.1943-7862.0000175.